\documentclass{article}

\usepackage{arxiv}

\usepackage[utf8]{inputenc} 
\usepackage[T1]{fontenc}    
\usepackage{hyperref}       
\usepackage{url}            
\usepackage{booktabs}       
\usepackage{amsfonts}       
\usepackage{nicefrac}       
\usepackage{microtype}      
\usepackage{lipsum}		
\usepackage{graphicx}
\usepackage[super, numbers, sort&compress]{natbib}
\usepackage{doi}
\usepackage{amsmath}
\usepackage{multirow}
\usepackage{caption}
\usepackage{subcaption}

\title{Energy Efficient Protein Language Models: Leveraging Small Language Models with LoRA for Controllable Protein Generation}

\author{ {Aayush Shah}\\
	Esperanto Technologies\\
	\texttt{aayush.shah@esperantotech.com} \\
	\And
	{Shankar Jayaratnam} \\
	Esperanto Technologies\\
	\texttt{shankar.jayaratnam@esperantotech.com} \\
}

\hypersetup{
pdftitle={A template for the arxiv style},
pdfsubject={q-bio.NC, q-bio.QM},
pdfauthor={David S.~Hippocampus, Elias D.~Striatum},
pdfkeywords={First keyword, Second keyword, More},
}

\begin{document}
\maketitle

\begin{abstract}
	Large language models (LLMs) have demonstrated significant success in natural language processing (NLP) tasks and have shown promising results in other domains such as protein sequence generation. However, there remain salient differences between LLMs used for NLP, which effectively handle multiple tasks and are available in small sizes, and protein language models that are often specialized for specific tasks and only exist in larger sizes.
In this work, we introduce two small protein language models, based on Llama-3-8B and Phi-3-mini, that are capable of both uncontrollable and controllable protein generation. For the uncontrollable generation task, our best model achieves an average pLDDT score of 69.75$\pm$12.74, demonstrating robust performance in generating viable protein structures. For the controllable generation task, in which the model generates proteins according to properties specified in the prompt, we achieve a remarkable average TM-Score of 0.84, indicating high structural similarity to target proteins. We chose 10 properties, including six classes of enzymes, to extend the capabilities of prior protein language models. 
Our approach utilizes the Low-Rank Adaptor (LoRA) technique, reducing trainable parameters to just 4\% of the original model size, lowering computational requirements. By using a subset of the UniRef50 dataset and small models, we reduced the overall training time by 70\% without compromising performance. Notably, Phi-3-mini reduced trainable parameters by 60\%, decreasing training cost by 30\% compared to Llama 3. Consequently, Phi-3 achieved a comparable TM-Score of 0.81, demonstrating that smaller models can match the performance of larger ones, like Llama 3.  We also demonstrate the deployment of our models on the energy efficient ET-SoC-1 chip, significantly improving the TPS/W by a factor of 3. The models are available at \href{https://huggingface.co/Esperanto/Protein-Llama-3-8B}{https://huggingface.co/Esperanto/Protein-Llama-3-8B} and \href{https://huggingface.co/Esperanto/Protein-Phi-3-mini}{https://huggingface.co/Esperanto/Protein-Phi-3-mini}, encouraging further research and development in the field of protein language models.

\end{abstract}


\begin{figure}
    \centering
    \includegraphics[width=0.75\linewidth]{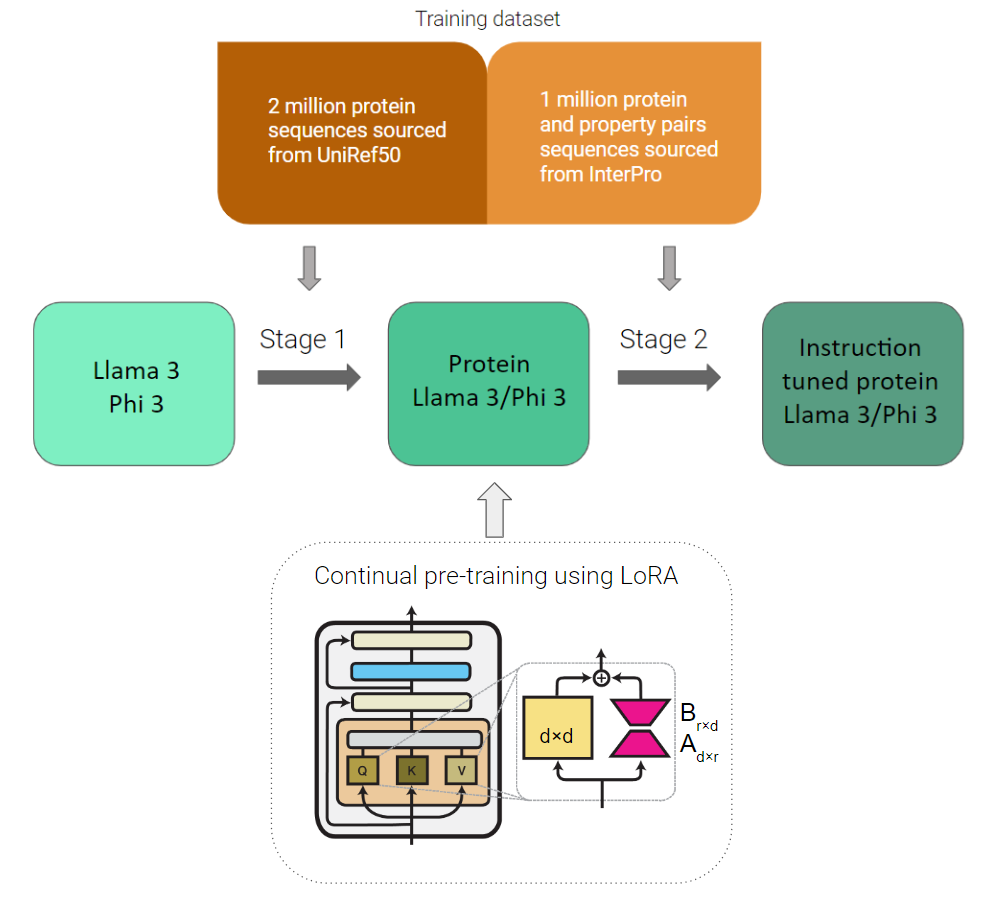}
    \caption{Training pipeline - Llama 3 and Phi 3 models were trained in two stages. First, they were trained on protein sequences and then on protein-property pairs. LoRA was used to lower the training cost.}
    \label{fig:train_pipe}
\end{figure}

\section{Introduction}

Large language models (LLMs) have revolutionized natural language processing (NLP) by demonstrating exceptional capabilities in a wide range of tasks, from text generation to translation and sentiment analysis \cite{naveed2024comprehensiveoverviewlargelanguage, zhang2023instructfingptfinancialsentimentanalysis, Zhao2023, zhu2024multilingualmachinetranslationlarge}. Models such as GPT-4 and Llama 3 have been termed as ‘foundational models’ due to their ability to generalize to multiple tasks by the process of fine-tuning on specialized datasets. \cite{achiam2023gpt, touvron2023llamaopenefficientfoundation}

Protein sequences are analogous to natural language because both are composed of a set of basic building blocks—amino acids for proteins and words for language—that combine to form meaningful structures. \cite{ofer2021language, yang2019machine, Valentini2023} In proteins, the sequence of amino acids determines the three-dimensional structure and function, whereas in natural language sentences, the sequence of letters and words determine the semantic meaning and local context. These similarities have prompted the application of large language models to various tasks like structure prediction from protein sequence \cite{Senior2020}, de-novo protein sequence generation \cite{Ferruz2022}, also termed as uncontrollable generation, and generation of proteins which possess certain properties or are specialized for a certain task, termed as controllable generation. Protein design finds its application in drug development and in the development of efficient artificial enzymes that break down industrial waste or plastics, contributing to carbon neutrality \cite{HUANG2023100446}. 

Recently there has been a drive towards the development of unified protein foundational models, similar to its counterparts in natural language, which possess an understanding of protein structure, sequence and they way they relate to each other. Such models have demonstrated the creation of novel proteins, effectively speeding up the evolutionary process \cite{Hayes2024.07.01.600583}. However, current protein foundational models are often extremely large in size and are generally closed source, which presents several challenges. Their vast size necessitates substantial computational resources for both training and inference, curbing their accessibility for many researchers and organizations. Additionally, the extensive computational power required for these large models leads to increased energy consumption, contributing to a higher environmental impact. 

As a result, there is a growing need for more efficient protein models that maintain high performance while being computationally economical. This could entail the usage of smaller language models, energy efficient hardware, or a combination of both. Lv et al. developed ProLLaMA \cite{lv2024prollama}, which demonstrated progress towards a unified protein foundational model by fine-tuning Llama 2 using LoRA \cite{hu2021loralowrankadaptationlarge} for the tasks of uncontrollable generation, controllable generation and property prediction. The usage of LoRA, which is parameter efficient fine-tuning technique, drastically reduces the training requirements and allows wide adoption of the model. 

However, it remains to be seen whether employing these fine-tuning techniques on newer state-of-the-art LLMs like Llama 3 and Phi 3 \cite{abdin2024phi} attain better performance with fewer training costs and energy consumption.

In this work, we introduce a two protein language models based on the Llama 3 and Phi 3 architectures. Using LoRA, we reduce the number of trainable parameters to just 4\% of the original model size, demonstrating the usefulness of our model in lowering the training cost. Considering the comparatively small size of Phi 3, we also reduce the training cost and inference time by imbuing Phi 3 with protein generation capability. Our major innovation is reducing the energy consumed during inference by 60\% by utilizing Esperanto Technologies’ energy efficient chip - ET-SoC-1 \cite{esperanto_ai}. ET-SoC-1 is based on the open-source RISC-V ISA and is designed for efficient, low-power computation. It is tailored for deploying generative AI models. We demonstrate the effectiveness of protein generation on ET-SoC-1 and prove that it is a viable alternative to GPUs for this task by assessing the quality of generated proteins. Following Lv et al., we utilize a two stage training process involving continual learning using LoRA on protein sequences for uncontrollable generation and subsequently fine-tuning it on an instruction dataset consisting of proteins linked with their properties for controllable generation.

A summary of our contributions is as follows:

\begin{itemize}
    \item Developed small protein language models based on the latest SOTA architectures, namely, Llama 3 and Phi 3.
    \item Demonstrated a 70\% reduction in the fine-tuning cost through the usage of LoRA and small language models.
    \item Extended the capabilities of protein language models to the task of controllable generation of enzymes.
    \item Reduced the energy consumption for inference by deploying the model on Esperanto's ET-SoC-1 chip. 
    \item Created a user-friendly web interface which displays key performance and inference metrics.

\end{itemize}

\section{Related Work}

\textbf{Sequence to structure: }The three-dimensional structure of a protein is linked to its function, making its determination essential for comprehending biological processes and understanding of medicine and life sciences. Among the landmark developments is AlphaFold, developed by Google DeepMind, which revolutionized the field with its unprecedented accuracy in predicting protein structures, as demonstrated in the CASP13 competition. \cite{10.1093/bioinformatics/btz422} AlphaFold uses a neural network based architecture that predicts the 3D coordinates of all non-hydrogen atoms for a given protein using the protein sequence and sequence homology. Subsequently, AlphaFold 2 and AlphaFold 3 have been developed, each extending the capabilities of its predecessor with improvements in the performance. \cite{Jumper2021, Abramson2024} AlphaFold 2 makes use of multiple sequence alignments (MSA) that identifies similar sequences that have been found in living organisms. AlphaFold 3, while being more accurate than its predecessor, makes predictions not just about the protein structure, but also the interactions between proteins and biological molecules such as DNA, RNA and ligands. RoseTTAFold is another structure prediction model which uses multiple neural networks to quickly and accurately predict protein structures based on amino acid sequences. \cite{doi:10.1126/science.abj8754} It utilizes one, two and three dimensional information to collectively reason about the relationship between a protein’s chemical parts and its folded structure. OmegaFold predicts high-resolution protein structure from a single primary sequence alone through a geometry-inspired transformer model trained on protein structures. \cite{Wu2022.07.21.500999} ESMFold, from the Fundamental AI Research team at Meta (FAIR), uses a large protein language model as a backbone and generates structure prediction using only one sequence as input by leveraging the internal representations of the language model, making the inference much faster than other state-of-the-art models. \cite{Lin2022.07.20.500902}

\textbf{Protein language modeling: } Natural language processing techniques have been modified to process protein language due to the similarities between protein sequences, which can be seen as concatenation of one letter representation of amino acids, and natural language. In the realm of building embedding based representations, ProteinBERT is pre-trained on protein sequences using a masked language modeling objective. It is able to generate embeddings from a protein sequence that captures its important biophysical properties. \cite{10.1093/bioinformatics/btac020} For the task of protein sequence generation through causal language modeling, ProtGPT2 is a decoder-only model based on the GPT-2 architecture which has learned the protein language by training on 50 million protein sequences. \cite{Ferruz2022} PeptideGPT generates proteins conditionally based on a user-defined property. \cite{shah2024peptidegptgenerativedesignpeptides} ProGen, from Salesforce, aims to generate novel protein sequences in controllable fashion using a transformer-based architecture trained on 280 million proteins. \cite{Madani2023, nijkamp2022progen2exploringboundariesprotein} ProteinMPNN uses a message passing neural network comprising of 3 encoder and 3 decoder layers and 128 hidden dimensions which predicts protein sequences in an autoregressive manner for a fixed backbone of interest. \cite{Dauparas2022}

\textbf{Unification of protein language models: } Recently, there has been a shift in research to combine multiple tasks pertaining to proteins in a single model. This allows the model to gain a deeper understanding of protein language rather than being specialized for a niche task. ProLLaMA combined uncontrollable generation, controllable generation and property prediction tasks by fine-tuning Llama 2 model on a dataset consisting of user instructions. ESM 3 is a 98 billion parameter model that has been trained on three modalities corresponding to proteins - sequence, structure and function, through masked language modeling objective. This makes the model highly flexible and allows for chain-of-thought prompting to generate new proteins. \cite{Hayes2024.07.01.600583} Taking this area of research forward, we aim to unify the task of controllable and uncontrollable generation via latest state-of-the-art LLMs, namely Llama 3 and Phi 3.

\section{Methods}


\subsection{Dataset preparation}

For the stage one of training, which aims to make the model generate proteins unconditionally, we sourced 2 million sequences from the 2024\_03 release of the UniRef50 dataset. \cite{suzek2015uniref} To demonstrate attainment of comparable performance with fewer training resources, we did not utilize the full dataset. For stage two of training, involving the instruction based fine-tuning, we chose a basket of 10 properties to illustrate the effectiveness of our model on controllable generation pertaining to these properties. The composition of the instruction dataset is shown in \ref{tab:enzyme_data}. We tackle the task of enzyme generation given the class of an enzyme, which has not been explored before. Six classes of enzymes have been considered, namely, Oxidoreductase, Lyase, Ligase, Transferase, Isomerase and Hydrolase. Creation of new enzymes is essential to medicine since engineered enzymes can be used in diagnostics, enzyme replacement therapies, and as potential treatments for diseases by targeting and breaking down harmful molecules. Moreover, new enzymes can help address environmental challenges by enabling the breakdown of pollutants and waste, contributing to a more sustainable future. The sequences pertaining to these enzyme classes were sourced from the ECPred40 dataset created by Buton et al. \cite{buton2023predicting} The remaining four classes were chosen as the S-adenosyl-L-methioninedependent methyltransferase superfamily (SAM-MT), the Tetratricopeptide-like helical domain superfamily (TPHD), the Thioredoxin-like superfamily (Trx), and the CheY-like superfamily (CheY). The performance on these classes have been reported by Lv et al., hence to establish a common ground for the purposes of comparison, we chose to include these four classes in our instruction dataset. These sequences have been taken from the instruction dataset open sourced by Lv et al.

\begin{table}[h!]
\caption{Protein classes distribution and sequence lengths}
    \centering
    \begin{tabular}{lrr}
        \toprule
        \textbf{Protein Type} & \textbf{Number of Sequences} & \textbf{Average Sequence Length} \\
        \midrule
        SAM-MT & 195,820 & 188 \\
        TrX & 129,686 & 159 \\
        TPHD & 141,034 & 170 \\
        CheY & 133,126 & 152 \\
        Oxidoreductase & 23,901 & 343 \\
        Transferase    & 65,899 & 336 \\
        Hydrolase      & 35,758 & 323 \\
        Lyase          & 18,550  & 325 \\
        Isomerase      & 12,151  & 342 \\
        Ligase         & 24,010 & 465 \\
        \midrule
        \textbf{Total} & \textbf{779,935} & {} \\
        \cmidrule(r){1-2}
    \end{tabular}
    
    \label{tab:enzyme_data}
\end{table}

\subsection{Training pipeline}

To endow a base LLM model with protein language understanding, it is necessary to train them on a dataset of protein sequences. However, completely re-initializing the parameters for pre-training faces two disadvantages: (i) the huge amount of computational resources required, and (ii) elimination of its prior understanding of natural language, which is pertinent for multi-task capabilities by understanding user input. To circumvent this, we employed continual learning using LoRa based on a similar approach by Lv et al. This reduces the number of trainable parameters and helps retain the prior natural language knowledge and allow the model to interpret user instructions. We employed this approach for Llama-3-8b and Phi-3-mini-4k-instruct models. The models were trained for the task of causal language modelling, where the loss function employed is cross-entropy between the output of the model and the ground truth, which is just the inputs shifted by one token to the left. This can be written as:

\[ \mathcal{L} = - \sum_{t=1}^{T} \log P(w_t \mid w_1, w_2, \ldots, w_{t-1}) \]


where \( \mathcal{L} \) is the loss, \( T \) is the total number of words in the sequence, \( P \) is the predicted probability of word \( w_t \), and \( w_1, w_2, \ldots, w_{t-1} \) are the preceding words in the sequence. For Llama 3, LoRA adaptors were added to $W_q, W_k, W_v, W_o, W_up, W_{gate}$ and $W_{down}$ weights whereas for Phi 3 they were added to the $W_q, W_k, W_v, W_o, W_{up}$ and $W_{down}$ weights. An overview of the training pipeline can be seen in Figure \ref{fig:train_pipe}. A learning rate of 5e-5 with a warmup ratio of 0.03 was used. The models were trained on a single NVIDIA A100 80GB GPU, with the per device batch size being 8 for Llama 3 and 16 for Phi 3. LoRA rank was set to 128 and alpha parameter to 256 for both stages of training.  

\subsection{Model Inference}

The generation process from a trained model is affected by several key parameters, including repetition penalty, top-p, top-k, and maximum sequence length. The repetition penalty discourages the model from repeatedly generating the same token, encouraging more diverse outputs. Top-p (nucleus sampling) ensures that the model selects tokens from the smallest possible set whose cumulative probability reaches at least $p$, thus choosing from a dynamically adjusted subset of likely candidates. Top-k restricts the next token selection to the top $k$ most probable tokens, focusing on the most likely continuations. The maximum sequence length defines the upper limit on the number of tokens the model can generate in a single sequence, thereby controlling the overall output length. The value of top $k$ was kept at 40, top $p$ at 0.9 and repetition penalty at 1.2 according to the best performing parameters found by Lv et al. The length of generation was approximately kept at the average sequence length of the respective class being generated to increase the alignment between the ground truth protein distribution and generated outputs. 

\begin{figure}
    \centering
    \includegraphics[width=0.8\linewidth]{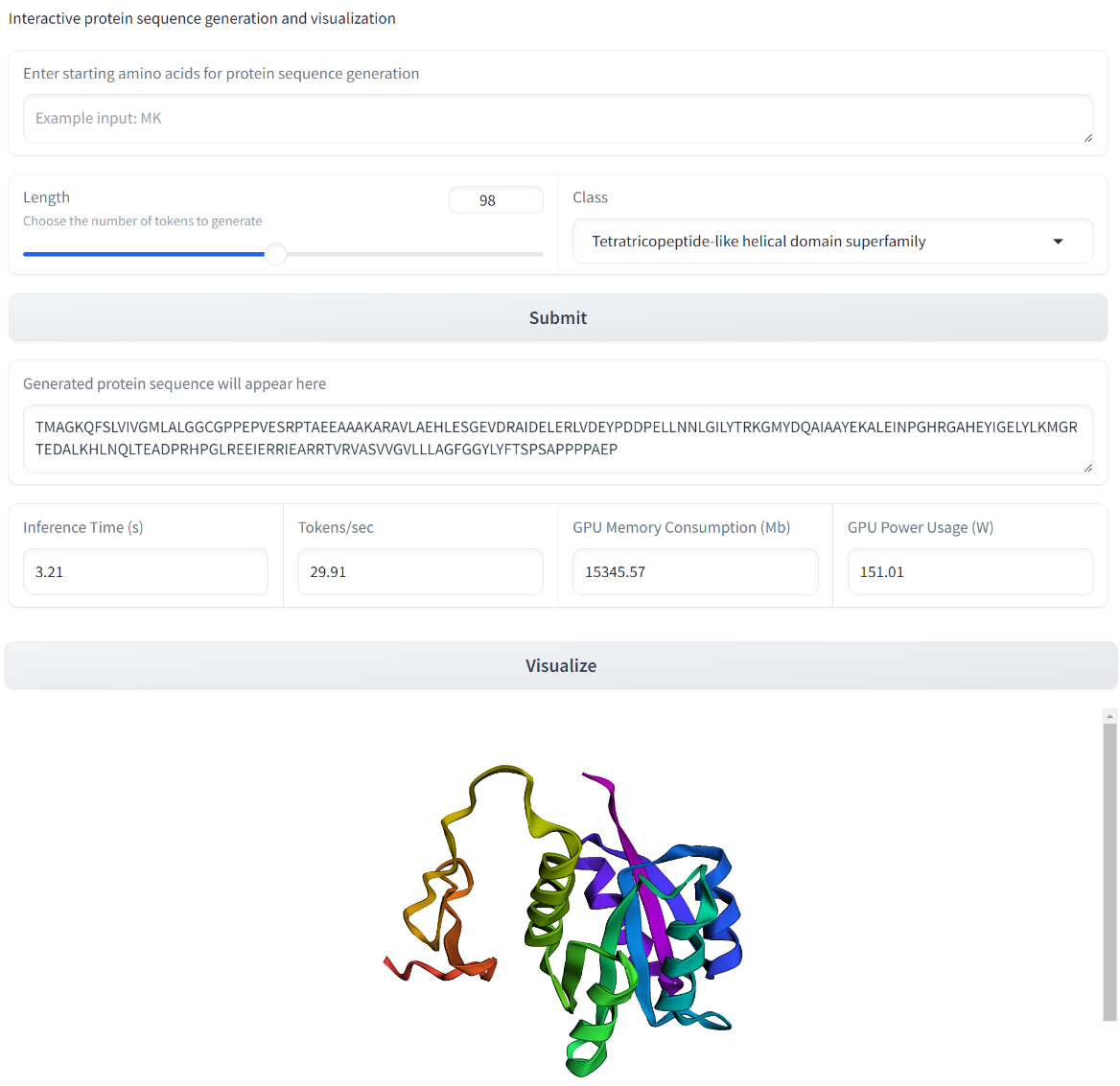}
    \vspace{2mm}
    \caption{Web interface for protein sequence generation and visualization. The user enters the starting amino acids, sequence length, and property for initiating the generation. The generated sequence and its structure shows up in the corresponding sections along with the performance metrics during inference.}
    \label{fig:enter-label}
\end{figure}

\subsection{Structure Evaluation}

Unlike natural language generated samples which can be evaluated on the basis of their grammatical correctness, coherence and relevance to the prompt by human evaluators, evaluation of protein sequences requires a dedicated model to predict the structural stability and quality of the generated proteins. This is quantified by the pLDDT score (predicted Local Distance Difference Test), which is a measure of the model's confidence in its structure prediction and has been known to correlate with orderness of the structure. We used ESMFold model to predict the structures and the pLDDT scores of generated proteins. ESMFold is a 15B parameter Transformer model and produces similar outputs compared to other state-of-the-art models like AlphaFold, with around 60 times faster inference. Regions with a pLDDT score between 60 and 90 are expected to be modeled accurately and exhibit an ordered structure, while lower scores (pLDDT less than 50) are typically found in disordered regions. 

\begin{figure}
    \centering
    \includegraphics[width=0.75\linewidth]{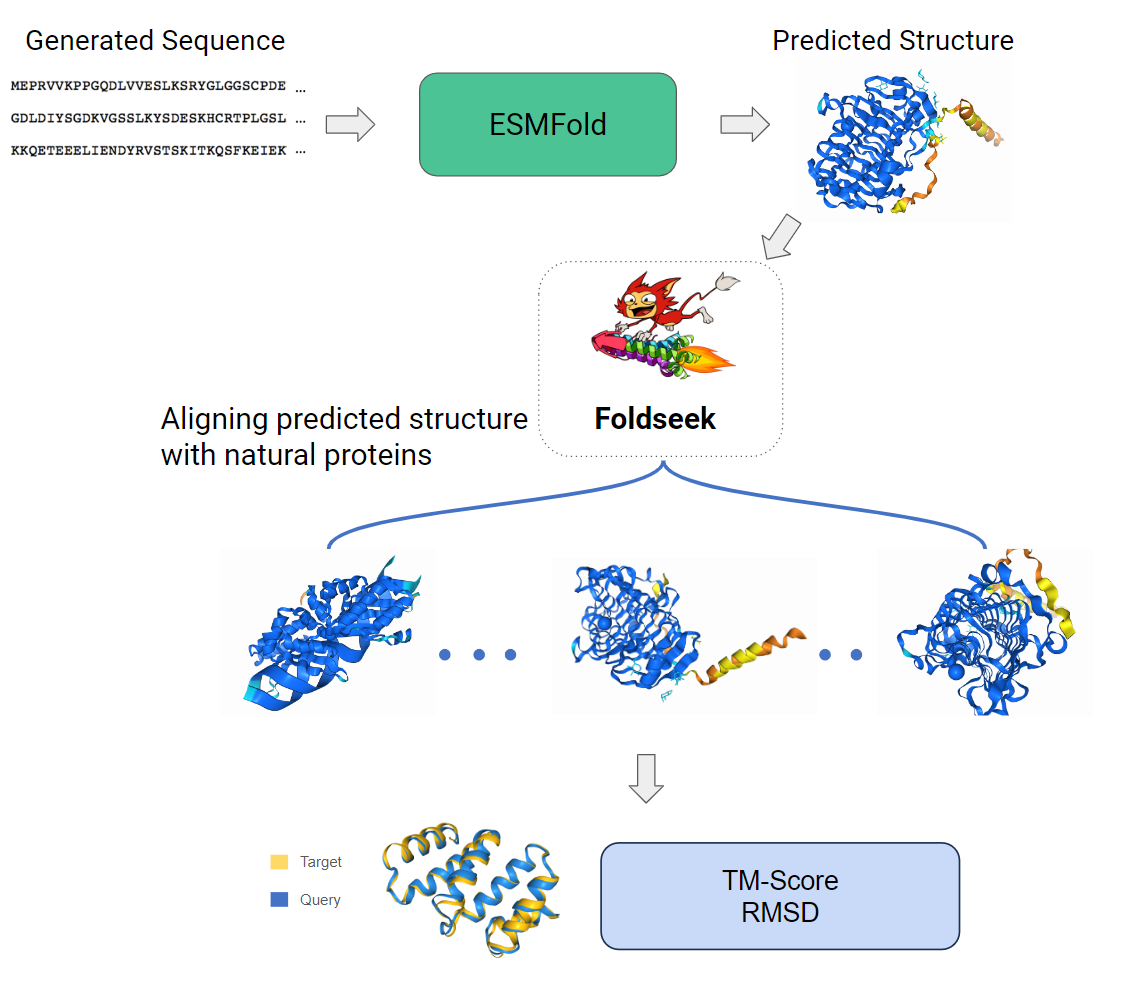}
    \caption{Evaluation pipeline - The generated sequence is passed to ESMFold for predicting its structure. Foldseek aligns this structure with known proteins and gives the TM-Score and RMSD value.}
    \label{fig:eval_pipe}
\end{figure}

\subsection{Controllable Generation}

To assess if the model is generating proteins according to the class mentioned in the input prompt, 100 sequences were generated per class. These were passed to ESMFold model to obtain their pLDDT scores, and sequences with a score lower than 60 were filtered out to ensure only sequences corresponding to stable structures were used in evaluation. The predicted protein structures were aligned with ground truth protein structures using Foldseek \cite{van2024fast}, and they were compared on the basis of TM-Score and homologous probability. TM-Score ranges from 0 to 1 and is a measure of the structural similarity between two protein structures, with higher scores indicating better alignment whereas homologous probability is the likelihood that two protein sequences share a common evolutionary ancestor. These scores tell us how similar a given protein is to a set of proteins belonging to a common class or property. 

\section{Results}

\begin{table}
    \caption{Performance comparison of selected methods.}
    \centering
    \begin{tabular}{llcccc}
        \toprule
        \multicolumn{1}{c}{\multirow{2}{*}{Method}} & \multicolumn{1}{c}{\multirow{2}{*}{pLDDT}} & \multicolumn{2}{c}{AFDB} & \multicolumn{2}{c}{PDB} \\
        \cmidrule(r){3-4} \cmidrule(r){5-6} 
        {} & {} & TM-score $\uparrow$ & RMSD $\downarrow$ & TM-score $\uparrow$ & RMSD $\downarrow$ \\
        \midrule
        ProtGPT2 & 56.32$\pm$16.05 & 0.44 & 12.60 & 0.43 & 9.19 \\
        ProGen2 & 61.07$\pm$18.45 & 0.43 & 15.52 & 0.44 & 11.02 \\
        ProLLaMA & 66.49$\pm$12.61 & 0.49 & 9.50 & \textbf{0.48} & 7.63 \\
        Llama 3 (A100) & \textbf{69.75$\pm$12.74} & \textbf{0.60} & 4.88 & 0.38 & 7.29 \\
        Llama 3 (ET-SoC) & 63.71 $\pm$ 6.78 & 0.45 & 6.49 & 0.34 & 4.11 \\
        Phi 3 (A100) & 60.41 $\pm$ 4.56 & 0.56 & \textbf{3.87} & \textbf{0.48} & \textbf{4.38} \\
        Phi 3 (ET-SoC) & 60.72 $\pm$ 6.58 & 0.48 & 7.14 & \textbf{0.48} & 4.62  \\
        \bottomrule
    \end{tabular}
    \label{tab:compare}
\end{table}

\subsection{Unconditional Generation}

The generation of sequences without providing the class of proteins as the context is termed as unconditional generation. We generated 100 sequences and evaluated them on the basis of their structural stability, as quantified through their pLDDT scores, and the measure of their alignment with two protein structure databases - AlphaFold Protein Structure Database (AFDB) and Protein Data Bank (PDB). The Swiss-Prot section of AFDB and the entire PDB was used from Foldseek to calculate the TM-Score and RMSD value in order to quantify the alignment. Table \ref{tab:compare} gives the results for our models compared with other state-of-the-art approaches. Our Llama 3 model obtains a pLDDT score of 69.75 $\pm$ 12.74, performing better than ProLLaMa's score of 66.49 $\pm$ 12.61. Llama 3 has a higher TM-Score and lower RMSD for AFDB database and a lower RMSD on the PDB database, while being sub-optimal on TM-Score. Phi 3 model obtained an average pLDDT score of 51.30 $\pm$ 6.90 but it secured higher RMSD scores for both AFDB and PDB. Considering the complexity of unconditional generation, our model has satisfactory performance and highlights the utility of LLM based methods in de-novo protein design.

\begin{figure}[htb]
    \centering
    \begin{subfigure}{0.24\textwidth}
        \centering
        \includegraphics[width=\textwidth]{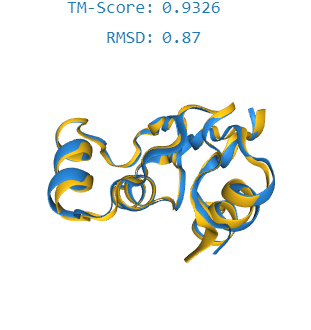}
        \caption{SAM-MT}
        \label{fig:fig1}
    \end{subfigure}
    \begin{subfigure}{0.24\textwidth}
        \centering
        \includegraphics[width=\textwidth]{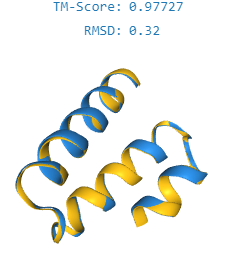}
        \caption{TPHD}
        \label{fig:fig2}
    \end{subfigure}
    \begin{subfigure}{0.24\textwidth}
        \centering
        \includegraphics[width=\textwidth]{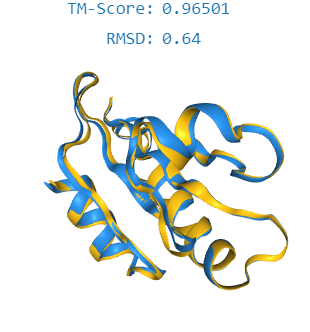}
        \caption{Trx}
        \label{fig:fig3}
    \end{subfigure}
    \begin{subfigure}{0.24\textwidth}
        \centering
        \includegraphics[width=\textwidth]{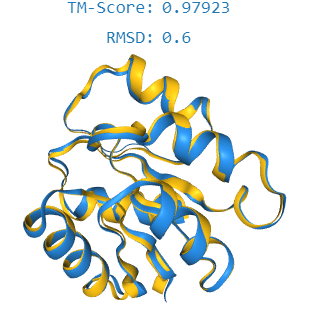}
        \caption{CheY}
        \label{fig:fig4}
    \end{subfigure}
    \caption{Structural alignment of proteins generated by Llama 3 for four different classes with their closest natural counterpart ranked according to TM-Score.}
    \label{fig:all_figures}
\end{figure}

\subsection{Controllable Generation}

Generating proteins according to some user-defined conditions is important to design proteins useful for different applications. We evaluated the performance of our models to generate proteins belonging to 10 different classes as highlighted in Table \ref{tab:controlled_results} on the basis of their TM-Score and RMSD with ground truth proteins. The high TM-Scores obtained by our models for these properties suggest that it is effectively capturing the underlying protein distribution pertaining to each of the classes. In the case of Oxidoreductase, the TM-Scores are below 75, implying sub-optimal modeling, however that can be ascribed to the inherent complexity associated with this class and the lesser data available in comparison. Llama 3 generally obtains higher score compared to Phi 3, with the exceptions being Transferase, Isomerase and Hydrolase, however the difference is not significant. This can be ascribed to the small size of Phi 3, which allowed it to be trained for a longer period. We also show the top structural alignments for four of the properties as given by Foldseek in Figure \ref{fig:all_figures}.

\begin{figure}
    \centering
    \includegraphics[width=0.75\linewidth]{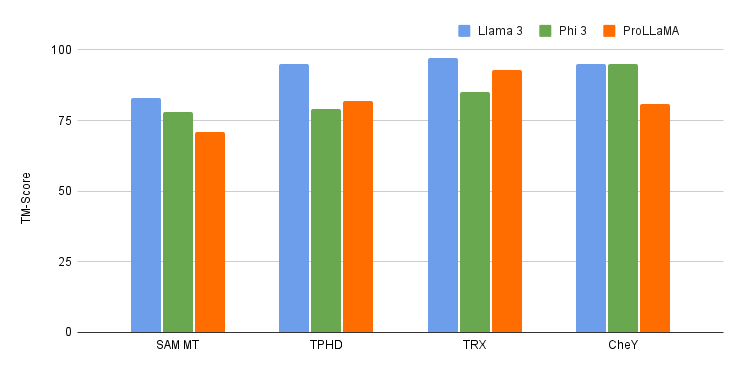}
    \caption{Comparison of best TM-Scores obtained by Llama 3 and Phi 3 on four properties}
    \label{fig:prop_compare}
\end{figure}

Comparing the performance of our model against ProLLaMA for the four properties in Figure \ref{fig:prop_compare}, we can see our model performs better against these benchmarks. This proves the capability of our approach in controllable generation and paves the way for integration of more properties.

\begin{table}
    \caption{Average TM-Score for 10 protein classes obtained by Llama 3 and Phi 3 on A100 GPU and ET-SoC-1}
    \vspace{2mm}
    \centering
    \begin{tabular}{ccccc}
        \toprule
        \multicolumn{1}{c}{\multirow{3}{*}{Class}} & \multicolumn{4}{c}{TM-Score} \\
        \cmidrule(r){2-5} 
        {} & \multicolumn{2}{c}{Llama 3} & \multicolumn{2}{c}{Phi 3} \\
        \cmidrule(r){2-3} \cmidrule(r){4-5}
        {} & A100 & ET-SoC-1 & A100 & ET-SoC-1 \\ 
        \midrule
        SAM-MT &\textbf{ 0.83} & 0.82 & 0.79 & 0.74 \\
        TPHD & \textbf{0.95} & 0.94 & 0.79 & 0.80 \\
        TRX & \textbf{0.97} & 0.84 & 0.86 & 0.85 \\
        CheY & \textbf{0.96} & 0.91 & 0.95 & 0.95 \\
        Ligase & \textbf{0.85} & \textbf{0.85} & 0.77 & 0.73 \\
        Hydrolase & 0.79 & 0.70 & 0.81 & \textbf{0.84} \\
        Lyase & \textbf{0.86} & 0.85 & 0.84 & 0.80 \\
        Oxidoreductase & 0.73 & \textbf{0.74} & 0.69 & 0.61 \\
        Transferase & 0.73 & 0.73 & \textbf{0.80} & 0.75 \\
        Isomerase & 0.81 & 0.84 & 0.86 & \textbf{0.88} \\
        \midrule
        \textbf{Average} & \textbf{0.84} & \textbf{0.82} & \textbf{0.81} & \textbf{0.81}\\
        \bottomrule
    \end{tabular}
    \label{tab:controlled_results}
\end{table}

\subsection{Inference metrics}

Measuring inference metrics such as tokens per second, energy consumption, and memory usage is crucial for evaluating the performance and efficiency of machine learning models, particularly in the context of large language models and other computationally intensive tasks. Tokens per second provides a measure of processing speed, indicating how quickly a model can generate or process text. Energy consumption reflects the power efficiency of hardware during inference, which is vital for assessing the sustainability and operational costs of deploying models at scale. Memory usage helps in understanding the resource requirements and limitations of the hardware, ensuring that models can be executed without running into memory bottlenecks.

Comparing different hardware platforms is essential for optimizing these metrics. Traditional GPUs, while powerful, can be energy-intensive and may not be the most efficient choice for all applications. In contrast, ET-SoC-1 offers a more energy-efficient alternative, potentially reducing power consumption while maintaining performance. ET-SoC-1 has an architectural design that focuses on minimizing energy usage and maximizing computational efficiency, which can lead to significant cost savings and lower environmental impact. By evaluating and comparing the performance of different hardware, developers and researchers can make informed decisions about the most suitable platform for their specific needs, balancing speed, efficiency, and operational costs.

We compare the performance during inference for Llama 3 and Phi 3 when deployed on NVIDIA A100 and ET-SoC-1 in Table \ref{tab:infer}. The models are in FP16 quantization, and we can see the significant reduction in power when for ET-SoC-1, dropping from 300 W in A100 to just 25 W in ET-SoC-1. The tokens-per-second also see a drop, however for TPS/W, a more energy conscious performance metric, we see a 60\% increase in Llama 3 and more than a three-fold increase for Phi 3. Hence, ET-SoC-1 consumes less energy than A100 GPU when generating the same number of tokens. These results highlight the importance of hardware for environmentally sustainable inference of protein language models.
\begin{table}
    \caption{Inference metrics comparison for NVIDIA A100 GPU and ET-SoC 1}
    
    \centering
    \begin{tabular}{lcccccc}
        \toprule
        \multicolumn{1}{c}{\multirow{2}{*}{Metric}} & \multicolumn{2}{c}{Llama 3} & \multicolumn{2}{c}{Phi 3} \\
        \cmidrule(r){2-3} \cmidrule(r){4-5} 
        {} & A100 & ET-SoC 1 & A100 & ET-SoC 1 \\ 
        \midrule
        Power (W) & 300 & {25} & 300 & {25} \\
        Tokens/sec (TPS) & 36 & {5} & 36 & {10} \\
        Memory (GB) & 15.35 & {15.35} & 7.36 & {7.36} \\
        TPS/W & 0.12 & \textbf{0.20} & 0.12 & \textbf{0.40}\\
        \bottomrule
    \end{tabular}
    \label{tab:infer}
\end{table}

To enable intuitive interaction with our models and subsequently visualize the structure of the generated protein sequences, we created a Gradio based user interface as shown in Figure \ref{fig:enter-label}. The user can input starting amino acids for generation, specify the length and the property, and get important inference metrics along with the generated output. 

\section{Conclusion}

In this work, we have introduced two compact protein language models, based on LLaMa-3-8B and Phi-3-mini, capable of both uncontrollable and controllable protein generation tasks. Our models demonstrate good performance in generating viable protein structures and synthesizing proteins with specific properties, effectively bridging the gap between large language models in natural language processing and specialized protein language models. For the uncontrollable generation task, our LLaMa-3-8B model achieved an average pLDDT value of 69.75±12.74, outperforming existing state-of-the-art models and showing robust performance comparable to naturally occurring proteins. This indicates our model's capability to generate high-quality protein sequences that align well with structural databases such as AFDB and PDB. We also explore the theme of energy efficient training and inference by reducing the number of trainable parameters through the Low-Rank Adaptation (LoRA) technique and improving the TPS/W by a factor of three for Phi 3 and 60\% for Llama 3 by deploying the model on the energy efficient ET-SoC-1 chip. Future work could involve expanding the range of controllable properties that the models can generate and including multiple modalities into the model like structure prediction, allowing for the design of proteins tailored to more specific and diverse biotechnological applications. Collaborations with experimental scientists to validate and refine the generated proteins in laboratory settings will be essential for translating computational models into tangible outcomes. Our work paves the way for the development of small protein language models and marks a significant step towards more sustainable AI practices in the field of protein language modeling.








\bibliographystyle{abbrvnat}
\bibliography{references}  






\section{Appendix}

\subsection{Training details}

\begin{figure}[h!]
    \centering
    \begin{minipage}[b]{0.45\textwidth}
        \centering
        \includegraphics[width=\textwidth]{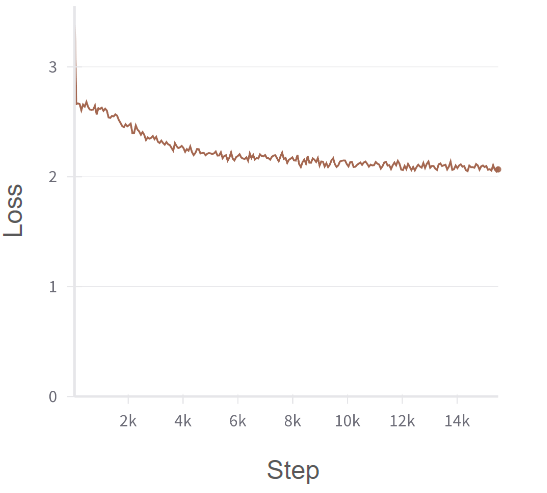}
        \subcaption{Llama 3 - Stage 1}
    \end{minipage}
    \hfill
    \begin{minipage}[b]{0.45\textwidth}
        \centering
        \includegraphics[width=\textwidth]{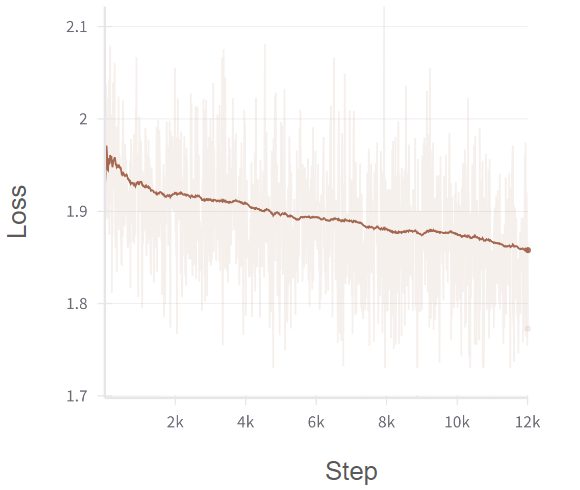}
        \subcaption{Llama 3 - Stage 2}
    \end{minipage}
    \vfill
    \begin{minipage}[b]{0.45\textwidth}
        \centering
        \includegraphics[width=\textwidth]{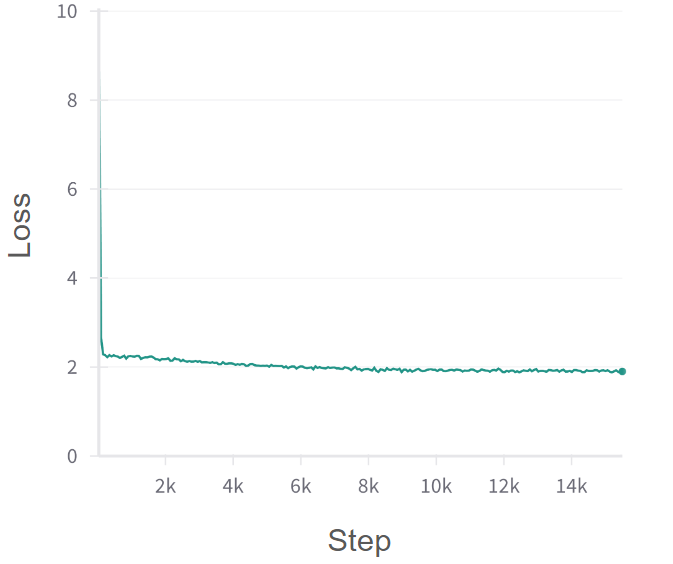}
        \subcaption{Phi 3 - Stage 1}
    \end{minipage}
    \hfill
    \begin{minipage}[b]{0.45\textwidth}
        \centering
        \includegraphics[width=\textwidth]{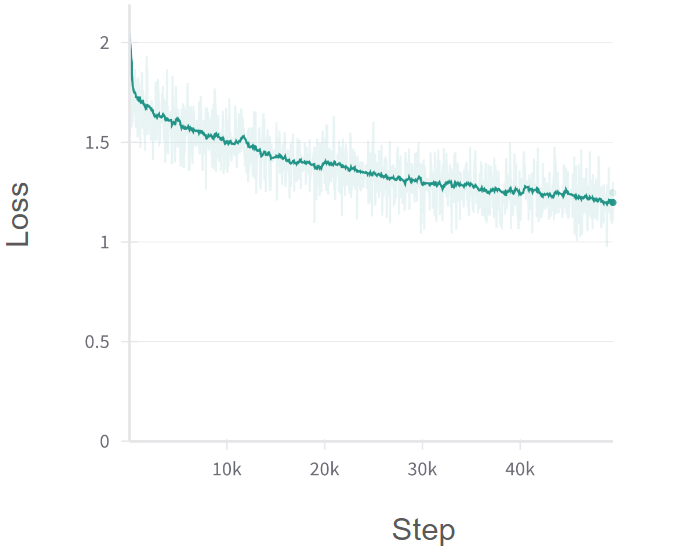}
        \subcaption{Phi 3 - Stage 2}
    \end{minipage}
    \caption{Training curves for Llama 3 and Phi 3 during both stages of training.}
    \label{train_curves}
\end{figure}

The models used in this work were Llama-3-8b and Phi-3-mini-4k-instruct downloaded from Hugging Face. For training, LoRA was employed with a rank of 128 and alpha parameter of 256. The rank and alpha were kept high due to the significant difference in the nature of natural language and protein language modeling. Learning rate was kept at 5e-5 for both the models. A batch size of 8 was used for Llama 3 and 16 for Phi 3. The models were trained on a single NVIDIA A100 80GB GPU. The training details are summarised in Table \ref{tab:train_params}. We can see the benefit of using Phi 3, a smaller language model compared to Llama 3, in reducing the training cost incurred. Phi 3 reduced the convergence time by 30\% during training, highlighting the importance of using small language models for the task of protein sequence generation. Figure \ref{train_curves} shows the loss convergence for Llama 3 and Phi 3 during stage one of training with losses not fully converging for stage two, indicating further room for improvement.

\begin{table}[h!]
\caption{Training parameters}
    \centering
    \begin{tabular}{lrr}
        \toprule
        \textbf{Training Parameter} & \textbf{Llama 3} & \textbf{Phi 3} \\
        \midrule
        Learning Rate & 5e-5 & 5e-5\\
        Batch Size & 8 & 16 \\
        Training Time - Stage 1 & 37 hours & 26 hours \\
        Training Time - Stage 2 & 48 hours & 35 hours \\
        LoRA rank & 128 & 128 \\
        LoRA alpha & 256 & 256 \\
        \hline
    \end{tabular}
    
    \label{tab:train_params}
\end{table}

\begin{table}[h!]
\caption{Generation parameters}
    \centering
    \begin{tabular}{lrr}
        \toprule
        \textbf{Generation Parameter} & \textbf{Llama 3} & \textbf{Phi 3} \\
        \midrule
        Top k & 40 & 40\\
        Top p & 0.9 & 0.9 \\
        Reptition penalty & 1.2 & 1.2 \\
        Max tokens - Unconditional & 70 & 30 \\
        Max tokens - Conditional & 100 & 100 \\
        \hline
    \end{tabular}
    
    \label{tab:gen_params}
\end{table}

\subsection{Generation Parameters}

We kept a top k, top p and repetition penalty of 40, 0.9 and 1.2 respectively since these parameters were reported to perform the best by Lv et al. For uncontrollable generation, we set the maximum number of generated tokens to 70 for Llama 3 and 30 for Phi 3, since the latter was the best at generating shorter sequences. On an average there are two amino acids per token, hence this corresponds to a protein length of 140 and 60 respectively. For controllable generation, since the ground truth proteins were themselves long, we set the number of generated tokens to 100 for both the models to facilitate maximum alignment with the ground truth proteins. A summary of generation parameters used can be found in Table \ref{tab:gen_params}.

\end{document}